%% file: main.tex
\documentclass[manuscript,screen,nonacm]{acmart}
\setcitestyle{numbers,sort&compress}
\usepackage{algorithmic}
\usepackage{textcomp}
\usepackage{booktabs,nicematrix,makecell}
\usepackage{xparse}
\usepackage{multirow}
\usepackage{lscape}
\usepackage{pifont}
\usepackage{tcolorbox}
\usepackage{xspace}
\usepackage{enumitem}
\usepackage{tabularx}
\usepackage{colortbl}
\usepackage{array}
\usepackage{wasysym}
\usepackage{cleveref}
\usepackage{acronym}
\usepackage{threeparttable}
\usepackage{float}
\usepackage{wrapfig}
\usepackage{algorithm}

\newcommand{\method}{\textsc{SPRINT}\xspace}

\newcommand{\std}[1]{\textsuperscript{\textcolor{gray}{\tiny$\pm$#1}}}
\newcommand{\algmeta}[1]{\par\noindent\begin{minipage}{\linewidth}\raggedright #1\end{minipage}\par}
\def\BibTeX{{\rm B\kern-.05em{\sc i\kern-.025em b}\kern-.08em
    T\kern-.1667em\lower.7ex\hbox{E}\kern-.125emX}}
\begin{document}

\title{\method: Robust Model Attribution of Generated Images via Secret Pixel Reconstruction}

\author{Kai Yao}
\email{kai.yao@ed.ac.uk}
\affiliation{%
  \institution{School of Informatics, University of Edinburgh}
  \city{Edinburgh}
  \country{United Kingdom}
}

\author{Marc Juarez}
\email{marc.juarez@ed.ac.uk}
\affiliation{%
  \institution{School of Informatics, University of Edinburgh}
  \city{Edinburgh}
  \country{United Kingdom}
}

\begin{abstract}
Detecting the source model of AI-generated images is a growing accountability problem. 
AI fingerprinting techniques address this by detecting imperceptible patterns in the images that are unique to each model, achieving high detection accuracy under ideal conditions.
However, recent research has shown that image fingerprints are extremely brittle to adaptive attacks, where knowledge of the technique can be exploited to perturb the fingerprints and evade detection.
We present \textbf{\method} (\textbf{S}ecret \textbf{P}ixel \textbf{R}econstruction fingerpr\textbf{INT}ing), a novel model attribution method specifically designed to provide robustness to adaptive attacks. As opposed to existing fingerprinting, which focuses on publicly discoverable patterns in the image, \method relies on a secret to define hidden reconstruction targets, thus keeping the verification task itself private.
As a result, the attacker can no longer see the task that the verifier solves at verification time, protecting the information exploited by the attacks.
Our results show that \method achieves high closed-world accuracy while remaining robust to adaptive attacks: on the FFHQ dataset, \method reaches $99.17\%$ clean accuracy on a diverse 12-model pool and $98.83\%$ on a harder pool of 6 close checkpoints of the same model architecture, while reducing adaptive removal and forgery attack success rates to $1\%$ or below. When the same pool of close model checkpoints is considered an open world, \method maintains high accuracy with an AUROC of $99.30\%$. These findings show that the approach of privatizing the verification task can make adaptive evasion substantially harder while maintaining performance in the clean setting.
\end{abstract}


\maketitle

\section{Introduction}

Image generation models are increasingly delivered as black-box services and integrated into downstream products. That creates an accountability problem; given a synthetic image and a set of candidate generators, how can an external verifier identify which generator produced the image without access to internals or the generation pipeline? We study this problem as \emph{model attribution}: identifying the source generator of a synthetic image. Source attribution matters for provenance, post-deployment oversight, and provider accountability, especially in high-stakes domains such as healthcare and defense~\cite{Bommasani2021,Terzis2024}. It also aligns with recent policy efforts that emphasize transparency, documentation, and traceability in AI systems~\cite{AIAct2023,WhiteHouseEO2023AI}.

A growing line of work studies \emph{passive fingerprints}, meaning statistical traces left by AI generators in their outputs. Passive traces support two related tasks: \emph{detection}, which separates synthetic from real images, and \emph{attribution}, which identifies the source generator~\cite{Nataraj2019CoOcc,Marra2019Fingerprints,mccloskey2018color,durall2020watch,dzanic2020fourier,frank2020leveraging,qian2020thinking,barni2020crossband,giudice2021dct,wang2020cnn,yu2019attributing,girish2021towards,nie2023attributing,Song2024ManiFPT,song2025riemannian,sha2023defake,li2024regeneration,wang2024latenttracer}. Passive attribution is attractive because it does not require modifying the generator, embedding a watermark, or controlling the generation service.
In standard \emph{closed-world} evaluations where all candidate generators are known, fingerprint-based passive attribution schemes can reach strong clean accuracy across architectures such as generative adversarial networks (GANs), variational autoencoders (VAEs), and diffusion models~\cite{yu2019attributing,girish2021towards,nie2023attributing,Song2024ManiFPT,song2025riemannian,sha2023defake,li2024regeneration,wang2024latenttracer}.

High clean attribution accuracy, however, is not enough for high-stakes applications where intentional evasion may cause serious harm. The security property we care about is whether the attribution scheme can be manipulated adaptively. Most existing passive attribution schemes expose a public deployed detector or public decision rule, or at least one that an attacker can learn well enough to attack. Their feature extractors, scoring rules, and decision procedures are fixed and published, or can be closely mimicked by surrogate models, which gives an adaptive attacker a direct target. Recent work shows that representative RGB, frequency-domain, and learned-feature schemes can be removed or forged under adaptive attack~\cite{wesselkamp2022misleading,yao2025smudged}. A passive attribution scheme can therefore look strong in a standard benchmark while offering little practical forensic value once the attacker knows what the verifier's detector is checking.

This paper revisits fingerprint-based passive attribution from that attack-surface perspective. Our goal is to keep attribution passive, i.e., not modifying or accessing the internals of the generator, while making the deployed detector depend on verifier-held private information. We propose \textbf{\method} (\textbf{S}ecret \textbf{P}ixel \textbf{R}econstruction fingerpr\textbf{INT}ing), a keyed passive attribution scheme. For each candidate generator, the verifier uses a verifier-held secret key to choose hidden pixel targets and trains a source-specific reconstructor to predict them. A reconstructor trained on one generator learns that source generator's regularities, so its hidden-target error remains low on that source generator and rises for others. In the typical closed-world setting, attribution then assigns an image to the enrolled generator whose source-specific reconstructor gives the lowest aggregate reconstruction error.

This construction keeps the practical appeal of passive attribution. Training requires only black-box access to generator outputs. The verifier does not modify generator parameters, alter the generation process, or require provider cooperation. What changes is the adaptive attack surface. Without the key, the attacker can no longer optimize against the same task that the verifier's deployed detector uses. Instead, the attacker must rely on surrogate detectors which, lacking access to the secret key, are likely to target incorrect pixels. In \method, a surrogate detector constructed with an incorrect key is misaligned in a strong sense: it is trained to predict different targets, so it solves a different task rather than approximating the actual verification task.

We first evaluate \method on classical closed-world attribution. The main experiments use two Flickr-Faces-HQ (FFHQ) benchmarks: a heterogeneous cross-family pool spanning GANs, VAEs, and diffusion models, and a harder pool of closely related StyleGAN2 checkpoints. We also report an open-world verification study on the same StyleGAN2 checkpoint pool motivated by checkpoint auditing. In that setting, a provider certifies one checkpoint and later serves a nearby, lower-cost checkpoint while still claiming that the audited checkpoint is in use. This is an open-world setting with single-source enrollment: the verifier trains only on outputs from the audited checkpoint and must later accept genuine samples while rejecting nearby substitutes that it never saw during enrollment. This reflects a third-party audit or compliance check for the case where a provider swaps the claimed checkpoint for a nearby substitute~\cite{matthias2004responsibility,Terzis2024}.

Our main finding is that keyed attribution preserves high clean utility, remains sharply sensitive to fine model changes in the near-checkpoint regime, and substantially reduces adaptive removal and forgery in the evaluated threat models. The same keyed design also supports a useful checkpoint-auditing capability through claimed-source verification: a detector enrolled from one audited checkpoint can remain sensitive to nearby substitutes without ever training on those substitutes as negatives. We also analyze how this capability behaves under realistic deployment factors such as base-rate shifts, partial secret-coordinate leakage, and comparison against representative public baseline schemes.

\smallskip
\noindent \textbf{Contributions.} In summary, our key contributions are as follows:
\begin{itemize}[leftmargin=25pt,topsep=2pt]
    \item We formalize the distinction between public and keyed passive attribution schemes, and we define the threat model where the attacker knows the scheme but not the verifier's secret key.
    \item We introduce \method, which derives secret pixel targets from a verifier-held secret key and trains one source-specific reconstructor per enrolled generator. In closed-world attribution, the detector ranks sources by reconstruction error; in open-world verification, the same detector construction supports single-source enrollment for checkpoint auditing.
    \item We evaluate \method on two FFHQ closed-world benchmarks. Clean accuracy reaches $99.17\%$ on a 12-generator cross-family pool and $98.83\%$ on a six-checkpoint near-checkpoint pool, while adaptive removal and forgery success rates stay around $1\%$ or below in both settings.
    \item We report a near-checkpoint claimed-source verification study motivated by checkpoint auditing. Across six audited checkpoints, single-source enrollment reaches mean AUROC of $99.30\%$ and rejects nearby substitute checkpoints that were never used as negatives during enrollment.
    \item We perform two ablation studies, showing: (i) \method outperforms a state-of-the-art fingerprinting scheme across varying amounts of secrecy in its fingerprints; (ii) \method's robustness drops sharply once a substantial fraction of the hidden task is exposed.
\end{itemize}

\section{Background and Related Work}

We introduce three parallel lines of work toward \emph{model attribution}. The first studies passive fingerprints for detection and attribution. The second studies active watermarking and private verification, which explains why public verification opens an attack surface. The third studies stronger cryptographic or systems-based assurance, which targets a different deployment model from ours.

\subsection{Passive Fingerprints for Synthetic Image Detection and Attribution}

Passive methods infer provenance from artifacts already present in the image, without modifying the generator or embedding an external signal. Early work focused on AI image detection: distinguishing GAN-generated images from real images using pixel-domain statistics, co-occurrence patterns, color features, frequency artifacts, and convolutional-neural-network (CNN)-based detectors~\cite{Nataraj2019CoOcc,Marra2019Fingerprints,mccloskey2018color,durall2020watch,dzanic2020fourier,frank2020leveraging,qian2020thinking,barni2020crossband,giudice2021dct,wang2020cnn,marra2020incremental}. Later studies extended these analyses beyond GANs to diffusion models and other generator families~\cite{Corvi2023Intriguing,corvi2023detection}.

A separate line of work studies source attribution: given a synthetic image, identify its source model. Existing passive attribution methods span noise-residual fingerprints~\cite{yu2019attributing}, open-world discovery of previously unseen GANs~\cite{girish2021towards}, latent fingerprints~\cite{nie2023attributing}, manifold- and geometry-based fingerprints~\cite{Song2024ManiFPT,song2025riemannian}, and text-to-image attribution methods based on discriminative classification, regeneration, or latent inversion~\cite{sha2023defake,li2024regeneration,wang2024latenttracer}.

Our work is closest to this passive attribution setting, but it differs in the security model. Rather than treating attribution as a fixed public prediction task, we introduce secrecy into the attribution system while remaining passive and black-box. This distinction is motivated by recent evidence that public fingerprint-based attribution pipelines can be highly vulnerable to adaptive removal and forgery attacks~\cite{wesselkamp2022misleading,yao2025smudged}.

\subsection{Active Watermarking and Private Verification}

Watermarking schemes modify training or generation to insert a recoverable signal~\cite{adi2018turning,chen2018deepmarks,boenisch2021systematic,yu2021artificial,Yu2023SelfWatermarked,lukas2023ptw,treering2023,gaussianshading2024}. Those approaches target a different deployment setting from ours because they require provider cooperation. They are nevertheless informative because recent attacks show that publicly verifiable watermarking schemes can also be attacked through surrogate keys or surrogate verifiers~\cite{lukas2023leveraging,fairoze2023publicly,fairoze2025difficulty,lin2025crack}. We draw the same high-level lesson for passive attribution: the key question is not only whether traces exist, but whether the verification function exposed to the attacker is public.

\subsection{Cryptographic and System-Level Assurance}

Cryptographic and systems-based approaches pursue a different form of assurance and authentication: instead of inferring provenance statistically from outputs, they try to attest that a particular model was \emph{executed} correctly, for example using verifiable computation, trusted execution environments, or zero-knowledge ML inference~\cite{ghodsi2017safetynets,tramer2018slalom,lee2024vcnn,zktorch,zkpytorch}. These approaches can in principle offer stronger guarantees based on cryptographic primitives, but they require substantially stronger deployment assumptions and often incur notable runtime overheads, which may render their deployment impractical in low-latency settings. Our goal is complementary: a lightweight passive attribution mechanism that can be trained from black-box access to a generator and later used by an external verifier without even provider cooperation.

\section{Problem Setting and Threat Model}
\label{sec:problem_setting}

This section defines the problem setting and threat model used throughout the rest of the paper. We first define passive source attribution and distinguish public attribution schemes from keyed, verifier-private schemes. We then decompose source attribution into enrollment and verification stages, and specify the two test-time settings considered in this work: closed-world attribution and open-world verification. Finally, we formalize the attacker capabilities, access levels, attack goals, and secret-compromise assumptions considered in our evaluation, and derive the design goals that motivate \method.

\subsection{Public and Keyed Passive Attribution}

We study source attribution for generated images. Let $\mathcal{G} = \{G_{\psi_1}, \dots, G_{\psi_n}\}$ be a pool of candidate source image generators with parameters $\psi_1, \dots, \psi_n$, each mapping latent and optional conditioning inputs to images in $\mathbb{R}^d$. Given a test image $x \in \mathbb{R}^d$, source attribution asks either to identify the generator that produced $x$ from a candidate pool, or to verify whether $x$ is consistent with a claimed source generator.

We call an attribution scheme \emph{passive} if it does not require the generator to embed an explicit watermark, signature, or provenance tag into its outputs. Instead, passive attribution relies on artifacts that arise naturally from the generation process, such as RGB-domain, frequency-domain, or learned feature statistics.

We use \emph{verifier} to refer to the party that constructs and deploys the attribution procedure, and \emph{detector} to refer to the concrete algorithmic instance produced by that procedure and used to score or classify test images. An attribution \emph{scheme} specifies how candidate generators are enrolled, what detector is produced, what information the verifier retains, and how test decisions are made.

We call a passive attribution scheme \emph{public} if its deployed detector can be fully reproduced from public information. Existing RGB, frequency, and learned-feature attribution schemes fall in this category~\cite{Nataraj2019CoOcc,dzanic2020fourier,giudice2021dct,wang2020cnn,yu2019attributing,Song2024ManiFPT,nie2023attributing,sha2023defake}. We call a scheme \emph{keyed}, or \emph{private}, if the deployed attribution detector depends on a verifier-held secret $K$ that is not publicly available. Under the same keyed scheme, different choices of $K$ instantiate different detector instances. Thus, the key distinction between public and keyed schemes is whether the exact deployed detector can be reproduced without compromising private information held by the verifier.

\subsection{Enrollment and Verification}

We decompose source attribution into two stages: enrollment and verification. Enrollment determines how the verifier constructs the detector and what information it retains before deployment, including any private state required by a keyed scheme. Verification determines how the deployed detector scores or classifies a test image. This separation lets us compare public and keyed attribution schemes under the same attribution pipeline while making explicit what information is available to the verifier and to the attacker.

\paragraph{Enrollment.}
Before deployment, the verifier enrolls one or more candidate generators using the chosen attribution scheme and prepares the resulting detector for test-time use. For notational simplicity, we use $\mathcal{G} = \{G_{\psi_1}, \dots, G_{\psi_n}\}$ to denote the enrolled generator pool. For a public scheme, all information needed to reproduce the deployed detector is public. For a keyed scheme, the deployed detector additionally depends on a verifier-held secret $K$. For open-world verification, the verifier also stores any decision threshold required by the scheme at the desired operating point.

\paragraph{Verification.}
At test time, we consider two settings. In \emph{closed-world attribution}, the detector assumes that $x$ was produced by one of the enrolled generators and returns the most likely source label in $\{1,\dots,n\}$. This setting compares $x$ against the full enrolled generator pool. In \emph{open-world verification}, the detector receives $x$ together with a claimed source $G_{\psi_t}$ and decides whether $x$ is consistent with that source. The detector checks only the claimed source and returns \texttt{Accept} or \texttt{Reject}. This setting is harder because the verifier may enroll only $G_{\psi_t}$, may not rely on alternative sources as negatives, and must set its threshold from genuine samples of $G_{\psi_t}$ alone. One motivation for this setting is \emph{checkpoint auditing} for regulatory purposes, where a model provider may claim to serve one checkpoint but instead deploy a cheaper substitute that harms user experience.

\subsection{Threat Model}

\paragraph{Attacker capabilities.}
We assume Kerckhoffs-style knowledge: the attacker knows the overall attribution scheme, including its implementation and enrollment procedure, as well as the candidate generators. For keyed schemes, however, the attacker does \emph{not} know the verifier's secret $K$.

In the closed-world robustness experiments, the attacker may use images from all generators in the enrolled pool to construct the attack, perturb test images before verification, and instantiate surrogate detectors when allowed by the access model. All attacks evaluated in this paper are \emph{offline with respect to the deployed detector}. The attacker may query the enrolled image generators to obtain images for surrogate training or attack construction, but does not query the deployed attribution detector during the attack. In particular, the attacker does not observe detector outputs for candidate perturbations and then adapt subsequent perturbations based on that feedback. This excludes interactive query attacks, such as repeatedly submitting modified images to the deployed detector, observing the predicted label, and using those responses to guide the next query.

\paragraph{Access levels.}
We distinguish access to the \emph{generator} from access to the \emph{detector}. On the generator side, black-box access means that the attacker can query a candidate generator and observe its outputs, but cannot inspect its weights or take gradients through it. White-box generator access would allow access to the generator architecture, weights, and gradients. In this paper, our attacks assume only black-box access to the generators and do not use generator architectures, weights, or gradients. This matches our focus on passive provenance, where attribution is performed from the image alone.

On the detector side, public and keyed schemes differ in what white-box access means. For a public scheme, white-box detector access means that the attacker can optimize directly against the exact deployed detector. For a keyed scheme, the attacker may know the scheme and may instantiate surrogate detectors using guessed or independently sampled secrets, but cannot optimize against the verifier's true deployed detector without knowing the true secret. We treat optimization against the true keyed detector as secret compromise rather than ordinary white-box access. Black-box detector access means that the attacker does not optimize against the deployed detector directly and must instead rely on transfer attacks or generic image transformation attacks.

\paragraph{Attack goals.}
We study two attack goals in the closed-world setting. Let $y$ denote the true source label of a test image $x$, and let $x'$ denote the perturbed image. A \emph{removal} attack succeeds when an image that was initially attributed to its true source is perturbed so that the detector predicts any other enrolled source, i.e., the prediction changes from $y$ to some label in $\{1,\dots,n\} \setminus \{y\}$. A \emph{forgery} attack succeeds when the perturbed image is attributed to an attacker-chosen wrong source.

In open-world verification, the analogous failures are false rejection and false acceptance. A false rejection occurs when a genuine image from the claimed source $G_{\psi_t}$ is rejected. A false acceptance occurs when an image from another source is accepted as consistent with the claimed source $G_{\psi_t}$. Our quantitative attack tables, \Cref{tab:cross-family-robustness,tab:near-checkpoint-robustness,tab:subset-ablation,tab:key-leakage}, report the closed-world version of these goals. The open-world experiments, \Cref{fig:claimed-source-case-study,tab:hardtarget-mxn-ablation}, instead study checkpoint auditing under provider-side substitution, where the provider swaps the claimed checkpoint for a nearby substitute; they do not evaluate active attacks on the detector.

\paragraph{Secret compromise.}
Our security claims are conditioned on the verifier's secret remaining unknown to the adversary. Operational key exposure, for example through social engineering or implementation compromise, is outside the scope of this paper. If a secret associated with a particular enrolled source is exposed, verification for that source is compromised. If a global secret used by the scheme is exposed, the full closed-world system is compromised. Our robustness target is therefore resistance to offline adaptive attacks without secret compromise. We separately study attacks that attempt to guess or recover the secret in \Cref{tab:key-leakage}.

\subsection{Design Goals}

The preceding problem setting and threat model motivate four design goals for a practical passive attribution scheme. First, the scheme should provide high clean utility in both closed-world attribution and the open-world checkpoint auditing setting considered in this work. Second, it should resist removal and forgery attacks under the offline, no-secret-compromise threat model. Third, enrollment should require only black-box access to the generator and should not require modifying generator parameters. Fourth, the verifier should remain independent of the provider after deployment, so that attribution can support auditing, dispute resolution, or other forms of third-party oversight.

\section{\method: Private Fingerprinting with Secret Pixel Reconstruction}
\label{sec:sprint-method}

This section introduces \method, a keyed passive attribution scheme that turns source attribution from public multi-class classification into a verifier-private pixel reconstruction (regression) task. Below, we describe enrollment with secret target coordinates, define the closed-world and open-world attribution decision rules, and explain the design intuition behind the method.

\subsection{Enrollment with Secret Pixel Targets}

\method is a keyed passive attribution scheme. Its detector is built around a simple idea: for each enrolled source, the verifier secretly chooses a small set of image coordinates and trains a source-specific reconstructor to predict the pixel values at those coordinates from the full image. A reconstructor trained on outputs from one generator should predict that source's secret targets better than it predicts targets on images from other sources. This gives the verifier a private source-specific score without modifying the generator or embedding a visible watermark.

This design also avoids a single public decision surface. Instead of training one global classifier shared by all sources, \method enrolls one source-specific reconstructor per generator. Each reconstructor depends on a secret coordinate set known only to the verifier.

Formally, let $K$ denote a master secret held only by the verifier. For each enrolled model identifier $\mathsf{id}_j$, the verifier first derives a per-source key using a key-derivation function (KDF):
\begin{equation}
    \kappa_j \gets \mathrm{KDF}(K, \mathsf{id}_j).
\end{equation}
The verifier then uses $\kappa_j$ to seed a deterministic coordinate sampler, denoted $\mathrm{DeriveIdx}$. In our empirical implementation, we use HMAC-SHA256 as the KDF, and the sampler expands the HMAC output with a standard seeded pseudorandom generator. Let $d$ be the image dimension; for an RGB image, $d = 3 \times \text{width} \times \text{height}$. Let $(l_1,\dots,l_H)$ be the head layout, with total hidden-target length $l=\sum_{h=1}^H l_h$. The sampler deterministically selects a pseudorandom subset of $l$ distinct flattened RGB scalar coordinates from $[d]=\{1,\dots,d\}$ without replacement, sorts the selected coordinates, and partitions the resulting ordered vector across the heads:
\begin{equation}
    S_j = (s_{j,1}, \dots, s_{j,H})
    \gets \mathrm{DeriveIdx}(\kappa_j, d, \{l_h\}_{h=1}^H).
\end{equation}
Each $s_{j,h} \in [d]^{l_h}$ is an ordered list of hidden scalar coordinates for head $h$. For an image $x$, we write $x_{S_j}$ for the ordered concatenation of $x_{s_{j,1}}, \dots, x_{s_{j,H}}$. These coordinates are not sampled once and stored as public metadata; only the verifier can regenerate them from its secret when needed. More details are explained in Appendix~\ref{sec:deriveidx_appendix}.

The learned part of the detector for source $j$ is a reconstructor $R_{\phi_j}$ with a shared convolutional backbone and $H$ regression heads, each with a structure such as a multilayer perceptron (MLP). Head $h$ outputs a vector in $\mathbb{R}^{l_h}$, and all heads together output $l$ scalar values. For an enrolled sample $x \sim G_{\psi_j}$, the target for head $h$ is
\begin{equation}
    y_{j,h}(x) = x_{s_{j,h}}.
\end{equation}
Training minimizes the average reconstruction error over heads:
\begin{equation}
    \mathcal{L}_j =
    \frac{1}{B}\sum_{i=1}^{B}
    \frac{1}{H}\sum_{h=1}^{H}
    \frac{1}{l_h}
    \left\|R_{\phi_j}^{(h)}(x_i) - y_{j,h}(x_i)\right\|_2^2.
\end{equation}
This reduces to the usual mean-squared error. The loss is applied only to the secret targets, and the detector never exposes the target coordinates or per-coordinate errors to the attacker. For open-world verification, the verifier also sets a decision threshold from genuine samples of the enrolled source for the chosen operating point.

\begin{figure*}[t]
  \centering
  \includegraphics[width=1.00\linewidth]{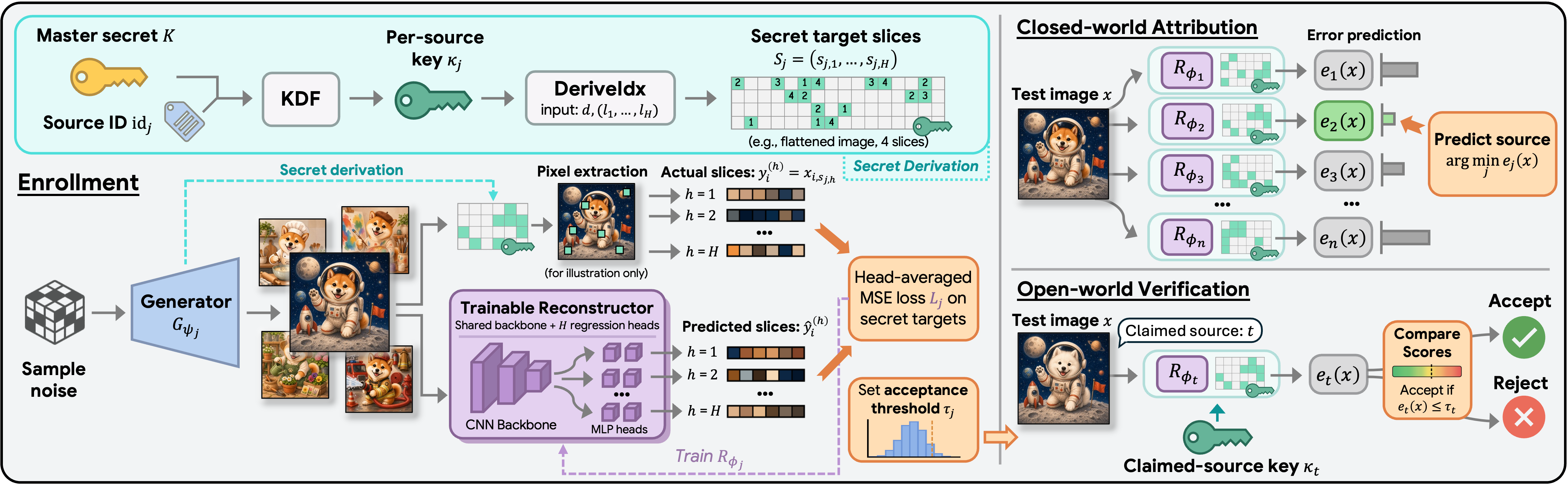}
  \caption{Overview of the \method pipeline. During enrollment, the verifier derives secret pixel targets from a private key, trains a source-specific reconstructor on generated samples, and sets thresholds for open-world verification. During verification, the verifier regenerates the same targets, compares predicted and observed pixel values at those locations, and uses the resulting reconstruction scores for closed-world attribution or open-world verification.}
  \Description{Pipeline diagram showing secret target selection, source-specific reconstructor training during enrollment, and keyed reconstruction-error scoring during verification.}
  \label{fig:sprint-pipeline}
\end{figure*}

\begin{algorithm}[htbp]
\caption{Enrollment for source $G_{\psi_j}$}
\label{alg:certification}
\algmeta{\textbf{Input}: master secret $K$, model identifier $\mathsf{id}_j$, input distribution $P_\mathrm{in}$, generator $G_{\psi_j}: \Xi \times \mathcal{C} \rightarrow \mathbb{R}^d$}
\algmeta{\textbf{Parameters}: head layout $(l_1,\dots,l_H)$, batch size $B$, learning rate $\eta$, training steps $T$}
\algmeta{\textbf{Output}: reconstructor $R_{\phi_j}$, private key $\kappa_j$, and open-world threshold $\tau_j$ when needed}
\begin{algorithmic}[1]
\STATE Derive per-source key: $\kappa_j \gets \mathrm{KDF}(K, \mathsf{id}_j)$
\STATE Derive secret coordinates: $S_j = (s_{j,1},\dots,s_{j,H}) \gets \mathrm{DeriveIdx}(\kappa_j, d, \{l_h\}_{h=1}^H)$
\STATE Initialize reconstructor $R_{\phi_j}$ with $H$ output heads of sizes $(l_1,\dots,l_H)$
\FOR{$t = 1$ to $T$}
    \STATE Sample batch $\{(\xi_i, c_i)\}_{i=1}^B \overset{\$}{\gets} P_\mathrm{in}$
    \STATE Generate images: $x_i \gets G_{\psi_j}(\xi_i, c_i)$
    \STATE Predict secret targets: $\hat{y}_i \gets R_{\phi_j}(x_i)$
    \STATE Extract target values: $y_i \gets \big(x_{i,s_{j,1}}, \dots, x_{i,s_{j,H}}\big)$
    \STATE Compute loss: $\mathcal{L} \gets \frac{1}{B}\sum_{i=1}^{B}\frac{1}{H}\sum_{h=1}^{H}\frac{1}{l_h}\|\hat{y}_i^{(h)} - y_i^{(h)}\|_2^2$
    \STATE Update reconstructor: $\phi_j \gets \phi_j - \eta\,\nabla_{\phi_j}\mathcal{L}$
\ENDFOR
\STATE Set open-world threshold $\tau_j$ from genuine samples of $G_{\psi_j}$ when open-world verification is used
\STATE \textbf{return} $R_{\phi_j}$ and $\tau_j$; keep $\kappa_j$ private
\end{algorithmic}
\end{algorithm}

\subsection{Closed-World Attribution}

For enrolled source $j$, the detector assigns test image $x$ the aggregate reconstruction error aggregated over all heads in a reconstructor
\begin{equation}
    e_j(x) =
    \frac{1}{H}\sum_{h=1}^{H}
    \frac{1}{l_h}
    \left\|R_{\phi_j}^{(h)}(x) - x_{s_{j,h}}\right\|_2^2.
\end{equation}
In closed-world attribution, the detector evaluates all enrolled sources and then predicts the source with the smallest reconstruction error:
\begin{equation}
    \hat{\jmath}(x) = \arg\min_{j \in \{1,\dots,n\}} e_j(x).
\end{equation}

\begin{algorithm}[htbp]
\caption{Closed-world attribution}
\label{alg:closed-world}
\algmeta{\textbf{Input}: test image $x$, enrolled reconstructors $\{R_{\phi_j}\}_{j=1}^n$, private keys $\{\kappa_j\}_{j=1}^n$}
\algmeta{\textbf{Output}: predicted source label $\hat{\jmath}(x)$}
\begin{algorithmic}[1]
\FOR{$j = 1$ to $n$}
    \STATE Regenerate $S_j$ from $\kappa_j$
    \STATE Compute aggregate reconstruction error $e_j(x)$
\ENDFOR
\STATE \textbf{return} $\arg\min_j e_j(x)$
\end{algorithmic}
\end{algorithm}

\subsection{Open-World Verification}
\label{sec:claimed-source}

In open-world verification, the detector checks one claimed source $G_{\psi_t}$. It evaluates only the reconstructor enrolled for that source. The detector averages the head errors into one score for the claimed source:
\begin{equation}
    e_t(x) =
    \frac{1}{H}\sum_{h=1}^{H}
    \frac{1}{l_h}
    \left\|R_{\phi_t}^{(h)}(x) - x_{s_{t,h}}\right\|_2^2.
\end{equation}
The detector accepts the claim iff $e_t(x) \le \tau_t$, where $\tau_t$ is the threshold set from genuine samples of $G_{\psi_t}$ for the chosen operating point.

\begin{algorithm}[htbp]
\caption{Open-world verification for a claimed source}
\label{alg:claimed-source}
\algmeta{\textbf{Input}: test image $x$, claimed source $t$, reconstructor $R_{\phi_t}$, threshold $\tau_t$, private key $\kappa_t$}
\algmeta{\textbf{Output}: verification decision}
\begin{algorithmic}[1]
\STATE Regenerate $S_t$ from $\kappa_t$
\STATE Compute score $e_t(x)$ for the claimed source
\STATE \textbf{return} \texttt{Accept} iff $e_t(x) \le \tau_t$
\end{algorithmic}
\end{algorithm}

\subsection{Design Intuition}

\paragraph{On why SPRINT can distinguish sources.}
Generators leave source-specific regularities in their outputs~\cite{theis2015generative,oord2016pixel,zong2018dagmm,ruff2018deepoc,gong2019memae}. \method turns those regularities into a keyed prediction task: the useful signal is not the secret pixel values by themselves, but how predictable those values are from the rest of the image. It is therefore useful to view the notion of ``fingerprint'' in \method not as the secret pixel values themselves, but as the source-specific pixel relationships that a trained reconstructor has learned in order to accurately predict those pixel value targets. A well-trained reconstructor, trained \textit{only} on images from one source, can use that source's regularities to predict the secret coordinates of an image with high accuracy, while the same prediction task becomes harder on images from other sources because they are out-of-distribution samples.

\paragraph{On using a small subset of pixels.}
The choice to predict a small secret subset of pixels but not a large subset is important. If the detector is designed to reconstruct the whole image or most of the image, the task would be largely public and easier for an attacker to approximate. Full-image reconstruction or reconstructing a large portion of image can also be dominated by generic semantic or perceptual structure, such as face layout or texture smoothness. These signals are useful for reconstruction quality per se, but they are not necessarily the source-specific traces needed for model attribution, at least not for models that generate sufficiently similar perceptual contents. In contrast, restricting predictions to a small subset forces the detector to tackle a localized prediction problem that exploits fine-grained dependencies among pixels in an image. In summary, the secret small subset makes the task hard to reproduce without the key, and the small subset design also keeps the score sensitive to subtle generator-specific residual structure rather than broad image semantics and perceptual contents, which is especially important in the open-world detection.

\paragraph{On the multi-head design.}
The multi-head design makes this prediction task less dependent on any single target slice and further increases detection sensitivity. The reconstructor uses a shared backbone, but each head predicts a different part of the secret target vector. The final score averages the head errors, so a test image must be consistent with several secret target groups rather than only one. This gives the detector multiple source-specific constraints that an input image must satisfy, enhancing open-world verification performance. We evaluate this design choice empirically in \Cref{tab:hardtarget-mxn-ablation} as an ablation experiment.

\section{Experimental Setup}
\label{sec:experiment_details}

This section describes how we instantiate and evaluate \method. We use the two verification settings introduced earlier. In \emph{closed-world attribution}, the detector compares a test image against all enrolled sources and predicts the source with the smallest aggregate reconstruction error. In \emph{open-world verification}, the detector checks a single claimed source and accepts the claim only when the claimed-source score is below a source-specific threshold.

Our experiments focus on both clean utility and robustness. The closed-world experiments evaluate whether \method can preserve attribution accuracy while resisting removal and forgery attacks. The open-world experiments evaluate checkpoint auditing utility: the verifier enrolls only the claimed checkpoint and must reject nearby substitutes without using them as negatives during enrollment or threshold setting.

\subsection{Evaluation Questions}

We organize the evaluation around five research questions (RQ):
\begin{enumerate}[label=\textbf{RQ\arabic*.}, leftmargin=*]
    \item Does \method preserve clean closed-world attribution accuracy while reducing adaptive removal and forgery attacks in a heterogeneous model pool? \(\rightarrow\) \Cref{subsec:cross-family-table}.
    \item Does the same robustness pattern hold when all candidate sources are closely related checkpoints? \(\rightarrow\) \Cref{subsec:near-checkpoint-table}.
    \item In the near-checkpoint setting, can \method support open-world checkpoint auditing with high utility, and which multi-head factorization works best? \(\rightarrow\) \Cref{subsec:open-world-auditing}.
    \item Is hiding a subset of coordinates from a public fingerprinting rule enough to provide robustness comparable to a keyed detector like \method? \(\rightarrow\) \Cref{subsec:coordinate-hiding}.
    \item How does the robustness of \method degrade when part of its secret coordinate set is leaked? \(\rightarrow\) \Cref{subsec:secret-leakage}.
\end{enumerate}

\subsection{Benchmarks and Verification Settings}

\paragraph{Closed-world attribution benchmarks.}
Closed-world results use two FFHQ benchmarks, both at $256 \times 256$ resolution.

\begin{itemize}[leftmargin=20pt,topsep=2pt]
    \item \textbf{Cross-family FFHQ pool.}
    We use the 12-model pool from a recent robustness benchmark~\cite{yao2025smudged}. It contains six GANs, namely R3GAN, GANformer, StyleSwin, CIPS, StyleGAN2, and StyleGAN3; three VAEs, namely VDVAE, NVAE, and VQ-VAE; and three diffusion models, namely ADM, NCSN++, and LDM. This benchmark measures attribution across architecturally diverse generator families.

    \item \textbf{Near-checkpoint FFHQ pool.}
    We also build a six-model pool from public FFHQ StyleGAN2 checkpoints released with the NVLabs StyleGAN2-ADA project~\cite{karras2020ada}. The pool crosses two training-set sizes, 30k and 70k FFHQ images, with three augmentation regimes: no augmentation, ADA, and ADA+BCR. These candidates share the same architecture and training pipeline, and differ only in training-set size and augmentation. This benchmark isolates the finer-grained setting where the candidate sources are very close substitutes.
\end{itemize}

The cross-family diverse pool tests broad source attribution. The near-checkpoint pool tests whether a detector can still distinguish sources that are intentionally much more similar perceptually.

\paragraph{Open-world checkpoint auditing.}
Open-world verification uses the same six-checkpoint StyleGAN2 pool. For each checkpoint $G_{\psi_t}$, we enroll one detector using only genuine samples from that checkpoint. The remaining five checkpoints are not used during enrollment or threshold setting; they are used only at test time as substitute sources. Thus, the open-world experiment consists of six independent single-source verification tasks. For each task, positives are samples from the audited checkpoint $G_{\psi_t}$, and negatives are samples from the other five checkpoints. \Cref{fig:claimed-source-case-study}(a) presents these six single-source detectors as a pooled six-by-six confusion matrix. \Cref{fig:claimed-source-case-study}(b) and \Cref{fig:claimed-source-case-study}(c) summarize the same six detectors with pooled ROC and precision--recall views.

\subsection{Instantiating \method}

\paragraph{Secret targets.}
All \method experiments instantiate the key-derivation function with HMAC-SHA256 under the verifier's master secret $K$. For each enrolled source identifier $\mathsf{id}_j$, the verifier derives a per-source key $\kappa_j$ and then samples the source's secret coordinate set $S_j$ from the flattened image domain. Coordinates are sampled without replacement from $[d]$, where $d = 3 \times 256 \times 256$ in our experiments. Unless noted otherwise, the main configuration uses total hidden-target length $l=32$ with $H=4$ heads of length $8$, i.e., $(l_1,l_2,l_3,l_4) = (8,8,8,8)$. The multi-head ablation in \Cref{tab:hardtarget-mxn-ablation} varies this layout while keeping the same general enrollment and scoring procedure.

\paragraph{Reconstructor architecture.}
Each enrolled source has its own reconstructor. The reconstructor uses a shared convolutional backbone followed by head-specific MLP regression heads. The backbone has six downsampling stages with channel widths
$[32,64,128,256,384,512]$. Each stage uses a $4 \times 4$ convolution with stride $2$ and padding $1$. We apply batch normalization after every stage except the first, followed by a LeakyReLU activation with negative slope $0.2$. Starting from a $256 \times 256$ RGB image, the backbone reduces the feature map to $4 \times 4$, and adaptive average pooling then produces a $1 \times 1$ feature tensor. Each head is a two-layer MLP with hidden widths $[256,128]$, LeakyReLU activations, dropout probability $0.3$, and a final linear layer that predicts that head's assigned target coordinates. In multi-head experiments, the head outputs are concatenated in the order induced by the secret coordinate set $S_j$. We keep the reconstructor architecture fixed and do not ablate it separately, since the architecture is an implementation choice and the paper focuses on the keyed verification design, target layout, and robustness evaluation under the stated threat model.

\paragraph{Training protocol.}
Reconstructors are trained with AdamW using learning rate $3 \times 10^{-4}$ and weight decay $10^{-4}$. The batch size is $32$ throughout. Samples are generated on the fly from the underlying generator at all three stages: training, threshold setting, and evaluation. This applies to \method and to all trainable baselines that we run ourselves. Unless noted otherwise, main benchmark results are averaged over five runs and reported as mean $\pm$ standard deviation.

\subsection{Baselines}

We compare SPRINT against seven representative public passive attribution baselines (Tables~\ref{tab:cross-family-robustness}~and~\ref{tab:near-checkpoint-robustness}). Nataraj19 uses RGB co-occurrence statistics~\cite{Nataraj2019CoOcc}. Giudice21 uses a low-dimensional discrete cosine transform representation~\cite{giudice2021dct}. Durall20 uses Fourier-spectrum discrepancies~\cite{durall2020watch}. Wang20 uses a learned CNN verifier~\cite{wang2020cnn}. Song24-RGB, Song24-Freq, and Song24-SL are the three public attribution variants reported in ManiFPT~\cite{Song2024ManiFPT}. For the public baselines in \Cref{tab:cross-family-robustness,tab:near-checkpoint-robustness}, we re-run clean and robustness evaluation on our benchmarks under the common attack protocol described below.

\subsection{Attack Protocols}

\paragraph{Scope.}
All attack-based robustness results in this paper are closed-world results. The attacker perturbs generated images directly. Once an image has been sampled, the generators are treated as black boxes: the attacks do not use generator weights or generator gradients. For the open-world checkpoint-auditing results in \Cref{fig:claimed-source-case-study} and the multi-head comparison in \Cref{tab:hardtarget-mxn-ablation}, the attack that we consider is a checkpoint substitution itself, not active attacks on the detector.

\paragraph{Attack families.}
We follow the attack categories of Yao and Juarez~\cite{yao2025smudged}: white-box attacks (WBA), surrogate attribution attacks (SAA), and generic transformation attacks (GTA). These categories refer to the attacker's access to the detector, not to the generator.

\begin{itemize}
    \item \textbf{WBA.}
    WBA is the strongest detector-aware attack we evaluate. For public schemes, WBA optimizes directly against the deployed detector, such as the published feature extractor, scorer, or learned classifier. When the fingerprint extraction algorithm defined in the scheme is differentiable, we directly apply gradient-based attacks. When it is non-differentiable, we train a surrogate extractor to approximate its fingerprint extraction behavior, then use the surrogate's gradients to guide the attack. For \method, the true deployed detector depends on the verifier's secret, so WBA cannot optimize through that detector without revealing the secret. Instead, the attacker instantiates surrogate detectors with guessed secrets, runs PGD against those surrogates, and measures success against the verifier's true detector. This is the keyed analogue of white-box optimization under the threat model in \Cref{sec:problem_setting}: the attacker knows the scheme but not the verifier's secret.

    \item \textbf{SAA.}
    In SAA, the attacker trains a separate attribution model from images alone as the surrogate. This surrogate model is trained for simple source classification from raw images only, without any knowledge on the fingerprint schemes. The attacker then attacks the true detector based on the gradients of this surrogate attribution model. 

    \item \textbf{GTA.}
    GTA applies generic image transformations such as JPEG compression, noising, blurring, and resizing, without requiring any information about the detector or even the generator pool. We evaluate GTA only as a removal attack because it is naturally an untargeted target.
\end{itemize}

\paragraph{Ablation-specific attack variants.}
Two later ablations reuse the same attack taxonomy. In \Cref{tab:subset-ablation}, \emph{Visible-only WBA} evaluates coordinate hiding for a public fingerprinting scheme. The attacker optimizes against only the visible part of the Giudice21 rule and does not observe the hidden coordinates. The SAA and GTA columns in that table keep the same meanings as in the main protocol. In \Cref{tab:key-leakage}, partial coordinate leakage is not a new attack family. It is the same \method WBA strategy, except that the attacker-side surrogate is given an increasing fraction of the true secret coordinates; leaked coordinates are inserted exactly, while the remaining coordinates are supplied by the guessed-key surrogate.

\paragraph{Success criteria and hyperparameters.}
A \emph{removal} attack succeeds when an image that was initially attributed to its true source is perturbed so that the detector predicts any other enrolled source. A \emph{forgery} attack is targeted: it succeeds when the perturbed image is attributed to an attacker-chosen wrong source. For forgery, each attacked image is assigned one target sampled uniformly from the non-source labels.

All attack success rates are computed only over the \emph{clean-correct} subset: for each generator, we first select up to 100 test images that are correctly attributed before any attack, and evaluate attacks only on those images. Clean images that are initially misattributed are excluded from the attack evaluation. PGD-style WBA and SAA use an $\ell_\infty$ budget of $0.05$, 50 steps, random start, and step sizes in $\{0.005, 0.01, 0.02, 0.05\}$. GTA uses Gaussian noise with standard deviation $0.005$, Gaussian blur with $\sigma=0.5$ and kernel size $3$, JPEG compression with quality $95$, and resizing by a factor of $0.9$.

\subsection{Metrics and Thresholds}

\paragraph{Closed-world attribution.}
For closed-world attribution, we report clean accuracy and attack success rate (ASR). Removal ASR measures the probability that an initially correct image is moved away from its true source. Forgery ASR measures the probability that an initially correct image is moved to the attacker's chosen target source. Closed-world decisions use the aggregate reconstruction errors $e_j(x)$ and predict the source with the smallest error.

\paragraph{Open-world verification.}
For open-world verification, we report ROC curves, precision--recall curves, average precision, AUROC, and fixed operating points denoted $q90$, $q95$, and $q99$. The detector uses the claimed-source score $e_t(x)$ from \Cref{sec:claimed-source}; lower scores indicate stronger consistency with the claimed source.

For a claimed source $t$, the threshold $\tau_t$ is set using positive samples from $G_{\psi_t}$ only. The $q90$, $q95$, and $q99$ thresholds are the empirical 90th, 95th, and 99th percentiles of positive scores, respectively. The reported TPR and FPR are then measured on the test set at these fixed thresholds, so the realized TPR may differ slightly from 90\%, 95\%, or 99\%. This positive-only thresholding matches the open-world checkpoint-auditing setting, where substitute checkpoints are unavailable during enrollment. As with any fixed threshold, the realized operating point can shift under distribution change.

\paragraph{Prevalence-adjusted precision--recall curves.}
The empirical open-world test set has a 1:5 ratio of positive to negative samples (as the pool has six models), but real-world deployment may have a different priors for the genuine model. We therefore also report precision--recall curves for varying prior distributions (i.e., base rates), which helps us assess real-world performance. For a threshold $\tau$, let $\operatorname{TPR}(\tau)$ and $\operatorname{FPR}(\tau)$ denote the measured true-positive and false-positive rates. Under deployment base rate $\pi \in (0,1)$, the corresponding PR point is $\bigl(\operatorname{TPR}(\tau), \operatorname{Precision}_{\pi}(\tau)\bigr)$, where $\operatorname{Precision}_{\pi}(\tau)=\pi\operatorname{TPR}(\tau)/\bigl(\pi\operatorname{TPR}(\tau)+(1-\pi)\operatorname{FPR}(\tau)\bigr)$. In \Cref{fig:claimed-source-case-study}, the highlighted empirical PR curve uses the actual 1:5 ratio, while the other PR curves show how precision changes under lower priors for the positive class.

\section{Results}
\label{sec:results}

We report results in the order of the research questions from \Cref{sec:experiment_details}. We first evaluate closed-world attribution on the cross-family pool and then on the harder near-checkpoint pool. We then use the same near-checkpoint pool for open-world checkpoint auditing and for the multi-head ablation study. Finally, we test two hypotheses that explain \method's effectiveness: whether hiding coordinates in a public fingerprint is enough, and how robustness of SPRINT changes when part of the secret coordinate set leaks.

\subsection{Cross-Family Closed-World Attribution}
\label{subsec:cross-family-table}

\input{tables/table_cross_family_robustness}

\Cref{tab:cross-family-robustness} answers RQ1. On the 12-model cross-family pool, \method (row 8) reaches $99.17\%$ clean accuracy and keeps WBA removal and forgery ASR at $0.54\%$ and $0.10\%$, respectively, showing that \method preserves clean utility while making adaptive removal and forgery considerably harder.

The public schemes (rows 1-7) expose a different attack surface. The best public WBA removal result is Song24-Freq at $31.59\%$, and the best public WBA forgery result is also Song24-Freq at $3.11\%$. Under SAA, the best public row is Song24-RGB, with $2.92\%$ removal and $0.44\%$ forgery. These are strong baselines, but none provides the utility--robustness tradeoff of \method. 
Recall that in the WBA column of \Cref{tab:cross-family-robustness}, the WBA strategy for public-schemes assumes full knowledge of the detector, while for \method, the adversary also has full knowledge of the detector but does not know the secret, and is therefore reduced to training surrogate models based on their guesses of the secret.

There are two more notable observations.
First, clean accuracy is not correlated with robustness. Wang20 reaches $98.38\%$ clean accuracy, close to \method, but its public detector is almost fully vulnerable under WBA, leading to $100\%$ removal and $99.28\%$ forgery ASR. Second, public schemes can be robust to certain attacks but not the others. Song24-RGB is comparatively strong under SAA and some generic transformations, but its clean accuracy is much lower and it remains vulnerable to WBA removal. Song24-Freq is the strongest public-WBA method, likely because the learned surrogate extractor cannot approximate the true fingerprint extraction process well enough; however, it still suffers from high SAA and resizing attacks in GTA. No public scheme is as robust across all attack families as SPRINT.

The GTA columns show that \method is not simply outperforming its public-scheme counterparts by being invariant to every image transformation. It stays below $1.3\%$ ASR for all four simple transformations, but Song24-RGB is even lower than SPRINT under Gaussian noise and JPEG compression. The takeaway on robustness from the results is relevant to security: \method does not claim to be immune to \textit{all} possible attacks, but it does remain competitive under simple corruptions while substantially improving robustness to WBA removal and forgery.

\subsection{Near-Checkpoint Closed-World Attribution}
\label{subsec:near-checkpoint-table}

\input{tables/table_near_checkpoint_robustness}

\Cref{tab:near-checkpoint-robustness} answers RQ2. This benchmark is harder because all six candidates are closely related StyleGAN2 checkpoints. Coarse generator-family cues are no longer available. Most public schemes suffer from a significant clean utility drop under this setting. However, \method still reaches $98.83\%$ clean accuracy and keeps WBA removal and forgery ASR at $0.83\%$ and $0.22\%$.

The public schemes again face a sharp utility--robustness tradeoff. Among public schemes (rows 1-7), the best WBA removal result is Song24-RGB at $41.76\%$, and the best WBA forgery result is Song24-Freq at $15.29\%$. Under SAA, \method reaches $0.94\%$ removal and $0.06\%$ forgery, while the best public scheme in SAA, Song24-RGB, reaches $3.06\%$ removal and $0.76\%$ forgery. Wang20 keeps high clean accuracy at $98.42\%$, outperforming other public schemes, but collapses under adaptive attacks, while Song24-RGB is remains robust against SAA and GTA attacks but only achieves $59.22\%$ clean accuracy.

This result provides more evidence about \method relying mainly on broad family-level differences. The detector is still able to separate checkpoints that share architecture and differ only in training-set size or augmentation. Under GTA, \method is best under noise, blur, and JPEG; under resizing, its ASR remains low at $2.35\%$. Overall, \Cref{tab:near-checkpoint-robustness} shows that the keyed reconstructor is effective at learning the keyed task even when the candidate sources are close substitutes.

This same pool also lets us ask a different question. Instead of asking which checkpoint produced an image, we ask whether an unseen new image is consistent with one claimed checkpoint in an open world of unknown checkpoints. We turn to this setting next.

\subsection{Open-World Checkpoint Auditing}
\label{subsec:open-world-auditing}

\input{figures/fig_open_world_case_study}

\Cref{fig:claimed-source-case-study} answers RQ3 in the open world setting. The audit question in this setting is: given a provider claiming to serve one checkpoint but who may instead actually use a nearby substitute, can \method verify the claim with high confidence? The verifier enrolls only the claimed checkpoint, sets a threshold from genuine samples of that checkpoint, and must reject samples from nearby checkpoints that were not used as negatives during enrollment.

\method remains highly accurate in this near-checkpoint setting. Across the six audited checkpoints, the mean AUROC is $99.30\% \pm 0.71\%$. At the fixed $q95$ threshold, the detector achieves average TPR $95.0\% \pm 0.9\%$ at FPR $2.2\% \pm 2.1\%$. At the fixed $q99$ threshold, recall rises to $99.0\% \pm 0.2\%$ at FPR $3.4\% \pm 3.1\%$.

\Cref{fig:claimed-source-case-study} (a) shows the six single-source detectors as a confusion matrix. Each row uses the threshold for a different audited checkpoint in the pool.The high contrast between values in the diagonal and other columns mean that, for all checkpoints, genuine samples are accepted and nearby substitutes are mostly rejected. The remaining outliers are concentrated in a small number of exceptionally harder checkpoint pairs, especially among checkpoints that may share the same augmentation methods and differ mainly in training-set size (e.g., 30k ADA+BCR vs 70k ADA+BCR). This is why we show the full matrix alongside pooled AUROC.
The scalar summary is high, but the matrix reveals where the audit problem may be harder.

Panels (b) and (c) plot the ROC and precision--recall curves for the same six detectors. The ROC curve shows that the detector remains in a low-error regime around the fixed $q95$ and $q99$ operating points. The precision--recall graph shows the empirical 1:5 base rate and also illustrates how precision changes when in low base rate deployment scenarios.

Given that the pool is composed of very similar models, we believe \method is likely to perform well in much larger open worlds with models that are not necessarily near checkpoints; we leave future work to confirm this generalization claim in larger open worlds.
In addition, we do not evaluate active adaptive attacks against the open-world verifier. In this setting, the malicious action is checkpoint substitution itself, rather than detector-targeted perturbation as in the closed-world robustness experiments. The closed-world results already test offline adaptive attacks across the same six checkpoints and show low vulnerability, so the open-world study focuses on whether one-model enrollment can detect nearby substitutions.

\input{tables/table_hardtarget_mxn_ablation}

\paragraph{Multi-head layout design ablation.}
\Cref{tab:hardtarget-mxn-ablation} answers the multi-head part of RQ3. It uses the same near-checkpoint open-world setting but varies the target/head layout.
When the total length of the secret is fixed at 32, splitting the target across heads helps: $4\times 8$ outperforms $2\times 16$ and $1\times 32$ at $q90$, $q95$, and $q99$, reaching $0.34\%$ FPR at $q90$.
When the per-head length is fixed at 8, moving from $2\times 8$ to $4\times 8$ improves all operating-point FPRs and AUROC, but $8\times 8$ does not improve further.

These results confirm the design intuition from \Cref{sec:sprint-method}: few heads make the keyed task less diverse. Too many heads make each head carry less stable signal. In this near-checkpoint open-world setting, $4\times 8$ is the best tested factorization. We treat this result as an empirically supported design choice within our experimental regime, rather than as a universal architectural principle.

\subsection{Robustness of Partially Public Schemes}
\label{subsec:coordinate-hiding}

\input{tables/table_public_subset_ablation}

\Cref{tab:subset-ablation} answers RQ4. The question is whether a public fingerprint can become robust if the verifier hides part of its representation. We test this with Giudice21 because its fingerprint design has a simple 63-dimensional DCT representation and it has a strong clean utility. Its partially secret variants keep the same feature family, scorer, and attack protocol, but expose fewer coordinates to the attacker.

The results show that hiding coordinates in a public fingerprint helps some attacks, but it does not recover the utility--robustness tradeoff achieved by \method. As we reduce the number of visible coordinates in Giudice21's fingerprint from 63 coordinates to 32, 16, 8, and 4, clean accuracy drops from $93.57\%$ to $77.09\%$, $60.72\%$, $43.18\%$, and $36.48\%$, respectively. Forgery becomes harder early: moving from 63 to 32 visible coordinates lowers Visible-only WBA forgery from $99.83\%$ to $14.89\%$. Removal, however, remains weak, with SAA and GTA success above $90\%$ at the same point.

Pushing visibility down to 8 or 4 coordinates eventually further lowers removal rates, but only after clean accuracy has collapsed below $45\%$. This is the negative control we need: simply hiding part of a public readout is not the same as making the detector depend on privately shifting the verifier's task. The latter is what gives \method its stronger utility--robustness profile.

\subsection{Partial Disclosure of \method's Secret}
\label{subsec:secret-leakage}

\input{tables/table_key_leakage}

\Cref{tab:key-leakage} answers RQ5. This ablation keeps the same WBA procedure for \method, but reveals part of the true secret coordinate set to the surrogate detector. The row $0/32$ is the fully secret setting used in the main robustness tables. The row $32/32$ exposes the full coordinate set to the attacker. Intermediate rows reveal some true coordinates and leave the rest guessed.

Small leakage has limited effect. Revealing 4 or 8 of the 32 secret coordinates increases removal ASR from $0.54\%$ to $0.92\%$ and $1.20\%$, and forgery ASR from $0.10\%$ to $0.14\%$ and $0.18\%$. There is an inflexion point when a larger part of the hidden task is exposed. At 16 leaked coordinates, removal ASR jumps to $16.42\%$ and forgery ASR to $3.18\%$. At 24 revealed coordinates, they reach $28.62\%$ and $5.50\%$. With full leakage, they rise to $68.90\%$ and $36.88\%$. This means that
\method does not require perfect secrecy of every coordinate to remain useful, but it does rely on keeping enough of the target set hidden. The degradation is nonlinear: limited leakage is tolerable, while substantial leakage makes surrogate detectors much better aligned with the true detector.

\section{Discussion and Conclusion}

\method changes the attack surface for passive attribution. Public schemes expose a detector that can be reproduced or learned by an attacker. \method instead makes the detector depend on secret target coordinates. In the evaluated closed-world benchmarks, this keeps clean accuracy high while sharply reducing removal and forgery under adaptive attacks. On the 12-model cross-family pool, \method reaches $99.17\%$ clean accuracy with WBA removal and forgery ASR of $0.54\%$ and $0.10\%$, respectively. On the harder near-checkpoint pool, it reaches $98.83\%$ clean accuracy with WBA removal and forgery ASR of $0.83\%$ and $0.22\%$, respectively. \method's reframing of passive attribution as a private reconstruction problem is effective at achieving robustness to adaptive attacks.
In the evaluated settings, this preserves the non-invasiveness of passive attribution while improving robustness to adaptive attacks and enabling single-source checkpoint auditing among nearby substitutes.

\paragraph{Security implications.}
The main security lesson is that a keyed scheme is most useful when it privately shifts the task the detector solves. Hiding only part of a public fingerprint can reduce some attack success rates, but it does not give the same combination of utility and robustness. By contrast, \method changes what the attacker must approximate: without the verifier's secret, the attacker can build surrogate detectors, but those surrogates solve a different prediction problem. This does not prove robustness against every possible surrogate strategy, but it explains why the evaluated adaptive attacks transfer poorly.

\paragraph{Open-world auditing.}
Our results also show strong open-world performance. In the near-checkpoint audit setting, the verifier enrolls one claimed checkpoint and does not train on nearby substitutes as negatives. Even so, the detector reaches a mean AUROC $99.30\%$, with FPR $2.2\%$ at $q95$ and $3.4\%$ at $q99$. This means that \method can be useful for auditing when a verifier needs to check whether a provider is still serving a claimed checkpoint. In addition, the confusion matrix shows nuance: some checkpoint pairs are harder than others. Given the similarity of the checkpoints in the open world, these results are promising, but future work could further explore how they generalize to other generator families.

\paragraph{Why the design matters.}
The ablations help explain where the gains come from. The multi-head study shows that performance is not just a matter of increasing target image coordinate length or reconstructor model capacity. A balanced target layout performs best in the near-checkpoint open-world setting, with $4\times 8$ strongest across $q90$, $q95$, and $q99$. The study of partially hiding a public scheme's fingerprint shows this approach is weaker than privately shifting the verification task (as SPRINT does). The leakage study shows the same mechanism from the other direction: robustness stays high under small leakage, but degrades significantly once enough of \method's secret coordinate set is exposed. Together, these results support a simple interpretation: \method works because the detector checks a private prediction task, and robustness weakens when that task becomes easier to reproduce.

\paragraph{Deployment fit and limitations.}
\method is designed for auditing a known set of generators, not for attributing every image on the internet. It works best when the verifier can query each candidate generator during enrollment, train one reconstructor for each generator, and keep the secret used to choose target coordinates private. This makes the method a good fit for settings such as attribution over a bounded model pool, checkpoint auditing, checking whether a provider has substituted one model for another, and other compliance workflows. It is less suitable for internet-scale open-world deployment, where the possible sources are unknown or very large.

Our experiments are currently limited to FFHQ-based face generators. We evaluate one diverse pool of 12 models and one pool of six nearby StyleGAN2 checkpoints. These results show that private reconstruction can separate sources well in a challenging domain. However, they do not show that the same behavior will necessarily hold for broader image domains or larger source pools. In addition, our open-world experiments focus on provider-side checkpoint substitution and do not test active attacks against the detector. We leave for future work evaluating \method on broader domains, stronger post-processing pipelines, larger generator pools, and adaptive open-world attacks.

\bibliographystyle{unsrtnat}
\bibliography{reference}

\appendix

\section{Open Science}

The source code and reproduction artifacts are being prepared for public release. The artifact contains the source code, experiment configuration files, unit and smoke tests, a Dockerfile, and reproduction guidance for the experiments reported in this paper. In particular, it includes the implementation of \method, the public baseline wrappers, closed-world robustness evaluation code, open-world verification code, attack implementations, plotting utilities, and the YAML configurations used to reproduce the reported settings.

\section{Ethical Considerations}
\label{sec:ethics}

This work studies provenance for generated images and attacks on attribution systems. The evaluation uses generated images and FFHQ-based generators or benchmarks; the manuscript does \textbf{not} report collection of new human-subject data or inference of sensitive attributes. Potential harms include over-reliance on attribution as proof of authenticity, misuse of attribution against benign users, and adversarial reuse of attack descriptions. The paper mitigates these risks by framing \method as model-consistency evidence rather than a universal authenticity oracle, by reporting base-rate dependence and key-leakage limits, and by restricting claims to the evaluated threat models. Deployments should combine \method with secure key management, logging, governance procedures, and independent provenance signals.

\section{Generative AI Usage}
\label{sec:genai}

A generative AI assistant (ChatGPT) was used for language polishing and grammar editing during manuscript revision. Codex was used for assisting coding and monitoring empirical experiments. The authors remain fully responsible for verifying all codes, results, claims and citations before submission.

\section{Deterministic Coordinate Selection}
\label{sec:deriveidx_appendix}

This appendix describes how \method selects the secret pixel coordinates used as reconstruction targets. In the main text, we denote this procedure by $\mathrm{DeriveIdx}$. The procedure is deterministic: once the verifier fixes the master secret $K$, the enrolled model identifier $\mathsf{id}_j$, the flattened image dimension $d$, and the head layout $(l_1,\dots,l_H)$, the same coordinate family $S_j$ can be regenerated whenever needed. Here, $\mathsf{id}_j$ is a stable identifier assigned by the verifier, such as a checkpoint name or another unique source label.

In our implementation, the verifier first uses HMAC-SHA256 as the KDF to derive a per-source key $\kappa_j$ from $K$ and $\mathsf{id}_j$. The verifier then uses $\kappa_j$ as the seed for a deterministic random number generator. The random number generator is used only to choose coordinate indices. Thus, the selection is reproducible: the same $\kappa_j$ always gives the same coordinates, while a different source identifier or verifier secret gives a different coordinate set.

The image is treated as a flattened vector of RGB scalar values. For an RGB image of size $\text{width}\times\text{height}$, the flattened dimension is
\[
d = 3 \times \text{width} \times \text{height}.
\]
A coordinate in $[d]=\{1,\dots,d\}$ refers to one scalar channel value, not to an entire RGB pixel. For example, the red, green, and blue values at the same spatial location correspond to three different scalar coordinates after flattening.

Given head lengths $(l_1,\dots,l_H)$, the procedure first selects
\[
l = \sum_{h=1}^{H} l_h
\]
distinct scalar coordinates without replacement. It then sorts the selected coordinates to obtain an ordered vector and partitions that vector into consecutive slices, one per head. This is the variant used in our main experiments: one per-source key determines one ordered target vector, and the heads receive disjoint slices of that vector.

\begin{algorithm}[t]
\caption{$\mathrm{DeriveIdx}$: secret coordinate selection}
\label{alg:deriveidx}
\algmeta{\textbf{Input}: per-source key $\kappa_j$, image dimension $d$, head lengths $(l_1,\dots,l_H)$}
\algmeta{\textbf{Output}: secret coordinate family $S_j=(s_{j,1},\dots,s_{j,H})$}
\begin{algorithmic}[1]
\STATE $l \gets \sum_{h=1}^{H} l_h$
\STATE Use $\kappa_j$ as the seed for a deterministic random number generator
\STATE Select $l$ distinct scalar coordinates $u_1,\dots,u_l$ from $[d]=\{1,\dots,d\}$ without replacement
\STATE Sort the selected coordinates to obtain an ordered vector $u=(u_{(1)},\dots,u_{(l)})$
\STATE $a \gets 1$
\FOR{$h = 1$ to $H$}
    \STATE $s_{j,h} \gets (u_{(a)},\dots,u_{(a+l_h-1)})$
    \STATE $a \gets a + l_h$
\ENDFOR
\STATE \textbf{return} $S_j=(s_{j,1},\dots,s_{j,H})$
\end{algorithmic}
\end{algorithm}

The output $S_j$ is a group of ordered coordinate lists. Each list $s_{j,h}$ has length $l_h$ and gives the target coordinates for head $h$. For an image $x$, the target vector $x_{S_j}$ is formed by extracting the scalar values at these coordinates and concatenating the head targets in order:
\[
x_{S_j}
=
\bigl(x_{s_{j,1}},\dots,x_{s_{j,H}}\bigr).
\]
This is the vector that the source-specific reconstructor is trained to predict.

Because the selection is without replacement, no scalar coordinate appears in two different heads for the same source. Across different sources, coordinate sets are generated from different source identifiers and may overlap by chance. This is not a problem because each source has its own coordinate set and its own reconstructor.

As a concrete example, suppose the verifier enrolls a checkpoint with identifier $\mathsf{id}_j=\texttt{ffhq70k-ada-bcr}$ under its master secret $K$. The KDF maps $(K,\mathsf{id}_j)$ to a per-source key $\kappa_j$, which is then used as the seed for coordinate selection. If the images are RGB images at $256\times256$ resolution, then the flattened image dimension is $d=3\times256\times256=196{,}608$. If the detector uses four heads with eight targets per head, then the head layout is $(l_1,l_2,l_3,l_4)=(8,8,8,8)$ and the total hidden-target length is $l=32$. With these inputs, $\mathrm{DeriveIdx}(\kappa_j,d,\{l_h\}_{h=1}^4)$ deterministically selects 32 distinct scalar coordinates from $[196{,}608]$, sorts them, and assigns eight coordinates to each head.

\end{document}

%% file: tables/table_cross_family_robustness.tex
\begin{table*}[t]
\centering
\small
\setlength{\tabcolsep}{2.1pt}
\renewcommand{\arraystretch}{1.08}
\begin{threeparttable}
\caption{Closed-world attribution on the 12-model FFHQ cross-family benchmark. The table reports clean accuracy and attack success rates. Entries are mean $\pm$ standard deviation over five runs. Rows are grouped by how WBA obtains gradients.}
\label{tab:cross-family-robustness}
\begin{tabular*}{\textwidth}{@{\extracolsep{\fill}}llccccccccc@{}}
\toprule
& & & \multicolumn{2}{c}{WBA ASR (\%)} & \multicolumn{2}{c}{SAA ASR (\%)} & \multicolumn{4}{c}{GTA Removal ASR (\%)} \\
\cmidrule(lr){4-5}\cmidrule(lr){6-7}\cmidrule(lr){8-11}
Method & Signal & Clean Accuracy (\%) & Removal & Forgery & Removal & Forgery & Noise & Blur & JPEG & Resize \\
\midrule
\multicolumn{11}{@{}l}{\emph{WBA through a learned surrogate extractor}} \\
Nataraj19~\cite{Nataraj2019CoOcc} & RGB & 91.98\std{1.35} & 91.14\std{8.07} & 8.92\std{3.32} & 91.39\std{2.24} & 10.86\std{0.92} & 26.14\std{1.08} & 37.50\std{3.24} & 48.67\std{4.32} & 56.83\std{2.35} \\
Song24-RGB~\cite{Song2024ManiFPT} & RGB & 64.24\std{0.87} & 95.36\std{2.25} & 12.56\std{4.04} & 2.92\std{0.38} & 0.44\std{0.08} & \textbf{0.03}\std{0.01} & 1.83\std{0.45} & \textbf{0.67}\std{0.28} & 0.20\std{0.04} \\
Durall20~\cite{durall2020watch} & Frequency & 80.11\std{1.26} & 95.93\std{6.19} & 58.83\std{5.14} & 99.48\std{0.45} & 18.57\std{1.32} & 18.11\std{5.28} & 60.31\std{2.24} & 38.01\std{5.17} & 93.11\std{1.36} \\
Song24-Freq~\cite{Song2024ManiFPT} & Frequency & 85.08\std{0.72} & 31.59\std{3.35} & 3.11\std{1.34} & 74.29\std{6.24} & 6.71\std{0.66} & 6.38\std{0.27} & 62.39\std{10.17} & 5.65\std{0.31} & 70.22\std{8.33} \\
Song24-SL~\cite{Song2024ManiFPT} & Learned & 85.10\std{0.93} & 85.36\std{1.10} & 10.00\std{1.38} & 62.19\std{2.08} & 7.97\std{1.34} & 1.78\std{0.19} & 16.61\std{1.16} & 6.34\std{0.28} & 37.38\std{2.20} \\
\midrule
\multicolumn{11}{@{}l}{\emph{Differentiable WBA}} \\
Giudice21~\cite{giudice2021dct} & Frequency & 93.57\std{1.33} & 100.00\std{0.00} & 99.83\std{0.38} & 91.67\std{7.23} & 9.75\std{5.20} & 32.03\std{1.32} & 65.92\std{5.38} & 44.25\std{5.14} & 87.25\std{1.31} \\
Wang20~\cite{wang2020cnn} & Learned & 98.38\std{0.35} & 100.00\std{0.00} & 99.28\std{0.74} & 89.83\std{5.19} & 6.08\std{2.07} & 0.81\std{0.38} & 39.58\std{7.26} & 35.50\std{6.21} & 48.42\std{5.34} \\
\method (Ours) & Learned & \textbf{99.17}\std{0.27} & \textbf{0.54}\textsuperscript{*}\std{0.10} & \textbf{0.10}\textsuperscript{*}\std{0.03} & \textbf{1.02}\std{0.24} & \textbf{0.19}\std{0.05} & 0.31\std{0.09} & \textbf{1.24}\std{0.18} & 1.18\std{0.11} & \textbf{0.09}\std{0.02} \\
\bottomrule
\end{tabular*}
\begin{tablenotes}[flushleft]
\footnotesize
\item Bold indicates the best value in each metric column. $^{*}$ For \method, WBA uses surrogate detectors with guessed secrets; the verifier's true secret is not revealed.
\end{tablenotes}
\end{threeparttable}
\end{table*}

%% file: tables/table_near_checkpoint_robustness.tex
\begin{table*}[t]
\centering
\small
\setlength{\tabcolsep}{2.0pt}
\renewcommand{\arraystretch}{1.08}
\begin{threeparttable}
\caption{Closed-world attribution on the FFHQ near-checkpoint benchmark, which contains six closely related StyleGAN2 checkpoints. The table reports clean accuracy and attack success rates. Entries are mean $\pm$ standard deviation over five runs. Rows are grouped by how WBA obtains gradients.}
\label{tab:near-checkpoint-robustness}
\begin{tabular*}{\textwidth}{@{\extracolsep{\fill}}llccccccccc@{}}
\toprule
& & & \multicolumn{2}{c}{WBA ASR (\%)} & \multicolumn{2}{c}{SAA ASR (\%)} & \multicolumn{4}{c}{GTA Removal ASR (\%)} \\
\cmidrule(lr){4-5}\cmidrule(lr){6-7}\cmidrule(lr){8-11}
Method & Signal & Clean Accuracy (\%) & Removal & Forgery & Removal & Forgery & Noise & Blur & JPEG & Resize \\
\midrule
\multicolumn{11}{@{}l}{\emph{WBA through a learned surrogate extractor}} \\
Nataraj19~\cite{Nataraj2019CoOcc} & RGB & 58.48\std{0.89} & 84.11\std{3.26} & 17.72\std{2.07} & 86.00\std{3.23} & 22.44\std{2.84} & 64.11\std{5.21} & 57.17\std{4.20} & 57.67\std{2.25} & 67.00\std{1.37} \\
Song24-RGB~\cite{Song2024ManiFPT} & RGB & 59.22\std{0.34} & 41.76\std{5.28} & 16.02\std{3.88} & 3.06\std{0.10} & 0.76\std{0.57} & 1.21\std{1.38} & 0.78\std{0.22} & 0.40\std{0.12} & \textbf{1.00}\std{0.27} \\
Durall20~\cite{durall2020watch} & Frequency & 78.15\std{1.08} & 92.29\std{2.11} & 26.63\std{1.17} & 80.90\std{4.04} & 18.92\std{1.24} & 16.80\std{3.19} & 69.97\std{5.08} & 25.62\std{2.04} & 71.90\std{3.10} \\
Song24-Freq~\cite{Song2024ManiFPT} & Frequency & 78.33\std{0.37} & 54.51\std{5.19} & 15.29\std{0.08} & 69.41\std{3.11} & 15.69\std{1.77} & 35.29\std{4.29} & 40.00\std{0.34} & 35.29\std{5.08} & 58.82\std{5.05} \\
Song24-SL~\cite{Song2024ManiFPT} & Learned & 57.30\std{1.38} & 76.49\std{4.37} & 19.90\std{4.26} & 74.50\std{3.31} & 19.11\std{2.30} & 5.28\std{3.24} & 26.83\std{2.22} & 15.50\std{1.02} & 38.67\std{3.09} \\
\midrule
\multicolumn{11}{@{}l}{\emph{Differentiable WBA}} \\
Giudice21~\cite{giudice2021dct} & Frequency & 62.95\std{1.73} & 100.00\std{0.00} & 99.94\std{0.02} & 85.78\std{1.13} & 24.17\std{1.33} & 14.06\std{2.36} & 63.33\std{3.24} & 63.50\std{4.35} & 78.00\std{5.05} \\
Wang20~\cite{wang2020cnn} & Learned & 98.42\std{0.22} & 100.00\std{0.00} & 100.00\std{0.00} & 85.50\std{2.23} & 54.39\std{3.17} & 7.61\std{2.36} & 8.67\std{0.38} & 12.00\std{1.37} & 6.17\std{2.87} \\
\method (Ours) & Learned & \textbf{98.83}\std{0.18} & \textbf{0.83}\textsuperscript{*}\std{0.22} & \textbf{0.22}\textsuperscript{*}\std{0.09} & \textbf{0.94}\std{0.16} & \textbf{0.06}\std{0.02} & \textbf{1.06}\std{0.30} & \textbf{0.07}\std{0.03} & \textbf{0.21}\std{0.15} & 2.35\std{0.23} \\
\bottomrule
\end{tabular*}
\begin{tablenotes}[flushleft]
\footnotesize
\item Bold indicates the best value in each metric column. $^{*}$ For \method, WBA uses surrogate detectors with guessed secrets; the verifier's true secret is not revealed.
\end{tablenotes}
\end{threeparttable}
\end{table*}

%% file: figures/fig_open_world_case_study.tex
\begin{figure*}[t]
\centering
\includegraphics[width=0.90\textwidth]{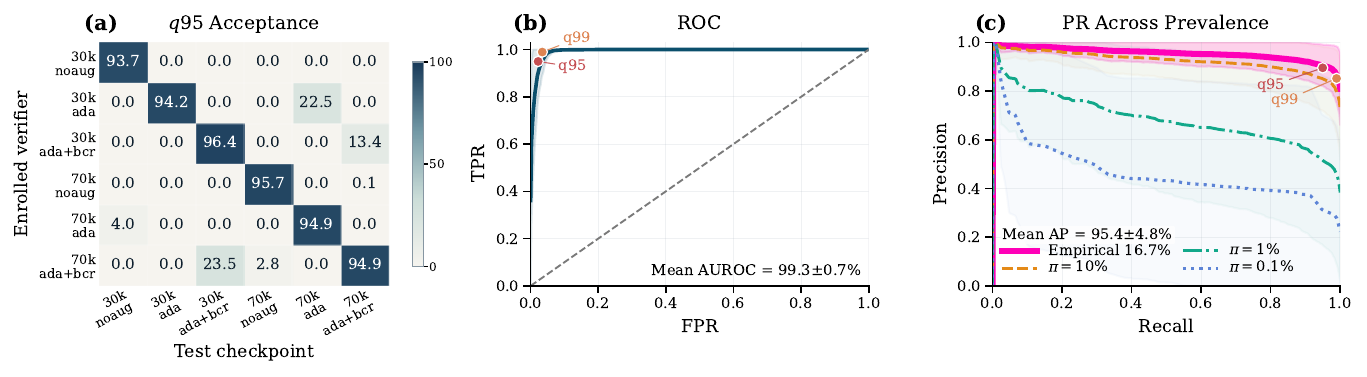}
\caption{Open-world verification for checkpoint auditing. The verifier enrolls only one audited checkpoint at a time and must reject nearby substitutes that were not used as negatives during enrollment or threshold setting. For each of the six checkpoints, we independently enroll one detector using genuine samples from that checkpoint only. Panel (a) stacks the six detectors into a pooled $q95$ six-by-six acceptance matrix, where each row is one audited checkpoint and each column is the checkpoint that produced the test image. Each row uses its own $q95$ threshold, set from genuine samples of that audited checkpoint. Thus, each entry reports the realized test-set acceptance rate under the fixed threshold for that row. Easier audited checkpoints remain close to diagonal, while the remaining substitute acceptance is concentrated in a small number of harder rows. Panel (b) shows the mean $\pm$ standard deviation ROC curve over the six audited checkpoints and marks the $q95$ and $q99$ operating points. Panel (c) shows precision--recall curves for the empirical one-versus-five evaluation prevalence and several lower deployment prevalences. The empirical prevalence, $\pi=16.7\%$, is highlighted, and the same $q95$ and $q99$ operating points are marked on that curve. Across the six audited checkpoints, \method reaches mean AUROC $99.30\%$.}
\Description{Three-panel open-world verification figure for checkpoint auditing. Panel (a) is a six-by-six q95 acceptance heatmap formed by stacking six independently enrolled detectors; each row corresponds to one audited checkpoint and each column to one test checkpoint. The matrix is close to diagonal for easier audited checkpoints, while a small number of harder rows show limited acceptance of nearby substitutes. Panel (b) shows the mean ROC curve over all six audited checkpoints with a shaded standard-deviation band and q95/q99 operating-point markers. Panel (c) shows precision--recall curves across prevalence levels, with the empirical one-versus-five evaluation prevalence emphasized, lower deployment prevalences shown alongside it, and q95/q99 marked on the empirical curve.}
\label{fig:claimed-source-case-study}
\end{figure*}

%% file: tables/table_hardtarget_mxn_ablation.tex
\begin{table}[t]
\centering
\small
\renewcommand{\arraystretch}{1.06}
\begin{threeparttable}
\caption{Head-layout ablation for open-world verification on the near-checkpoint benchmark. Results are averaged over the six audited checkpoints. The first block keeps the total number of secret targets fixed at 32 and changes how those targets are split across heads; the second block keeps each head at width 8 and changes the number of heads. The $q90$, $q95$, and $q99$ thresholds are set from genuine samples of the claimed checkpoint only.}
\label{tab:hardtarget-mxn-ablation}
\setlength{\tabcolsep}{4.0pt}
\begin{tabular*}{\columnwidth}{@{\extracolsep{\fill}}lcccc@{}}
\toprule
Head Layout & \shortstack{FPR at\\$q90$ (\%)} & \shortstack{FPR at\\$q95$ (\%)} & \shortstack{FPR at\\$q99$ (\%)} & AUROC (\%) \\
\midrule
\multicolumn{5}{@{}l}{\emph{Fixed total target length: 32}} \\
$1\times 32$ & 3.76 & 8.54 & 10.98 & 97.02 \\
$2\times 16$ & 1.23 & 3.10 & 4.05 & 98.92 \\
$4\times 8$ & \textbf{0.34} & \textbf{2.21} & \textbf{3.43} & \textbf{99.30} \\
\midrule
\multicolumn{5}{@{}l}{\emph{Fixed width per head: 8}} \\
$2\times 8$ & 1.54 & 3.06 & 4.05 & 99.03 \\
$4\times 8$ & \textbf{0.34} & \textbf{2.21} & \textbf{3.43} & \textbf{99.30} \\
$8\times 8$ & 1.60 & 3.26 & 4.01 & 98.86 \\
\bottomrule
\end{tabular*}
\begin{tablenotes}[flushleft]
\footnotesize
\item A layout $H\times l_h$ means $H$ heads, each predicting $l_h$ secret pixel targets. Bold indicates the best value within each ablation block.
\end{tablenotes}
\end{threeparttable}
\end{table}

%% file: tables/table_public_subset_ablation.tex
\begin{table*}[!t]
\centering
\small
\renewcommand{\arraystretch}{1.08}
\begin{threeparttable}
\caption{Coordinate hiding with the public Giudice21 scheme on the FFHQ cross-family benchmark. This ablation hides part of the public DCT representation rather than changing the detector task. In the WBA columns, the attacker optimizes only against the visible DCT coordinates; SAA and GTA follow the main closed-world attack protocol. Entries are mean $\pm$ standard deviation over five runs.}
\label{tab:subset-ablation}
\setlength{\tabcolsep}{4.2pt}
\begin{tabular*}{\textwidth}{@{\extracolsep{\fill}}lccccccc@{}}
\toprule
Variant & Visible DCT & Hidden DCT & \shortstack{Clean\\Accuracy (\%)} & \shortstack{WBA on Visible DCT\\Removal ASR (\%)} & \shortstack{WBA on Visible DCT\\Forgery ASR (\%)} & \shortstack{SAA\\Removal ASR (\%)} & \shortstack{GTA\\Removal ASR (\%)} \\
\midrule
Giudice21~\cite{giudice2021dct} & 63 & 0  & \textbf{93.57}\std{1.33} & 100.00\std{0.00} & 99.83\std{0.38} & 91.67\std{7.23} & 97.08\std{1.84} \\
Giudice21-L32 & 32 & 31 & 77.09\std{5.54} & 76.05\std{6.04}  & 14.89\std{1.04} & 91.61\std{3.18} & 93.11\std{2.27} \\
Giudice21-L16 & 16 & 47 & 60.72\std{3.82} & 80.36\std{2.72}  & \textbf{9.11}\std{1.92}  & 83.25\std{3.54} & 87.47\std{2.65} \\
Giudice21-L8  & 8  & 55 & 43.18\std{2.07} & 68.71\std{3.45}  & 12.63\std{2.14} & 69.89\std{4.22} & 51.95\std{5.12} \\
Giudice21-L4  & 4  & 59 & 36.48\std{4.21} & \textbf{40.09}\std{4.59}  & 9.35\std{2.08}  & \textbf{62.20}\std{4.75} & \textbf{36.37}\std{7.48} \\
\bottomrule
\end{tabular*}
\begin{tablenotes}[flushleft]
\footnotesize
\item Bold indicates the best value in each metric column. The variants Giudice21-L32/L16/L8/L4 expose only the first 32/16/8/4 DCT coordinates to the attacker and hide the remaining coordinates.
\end{tablenotes}
\end{threeparttable}
\end{table*}

%% file: tables/table_key_leakage.tex
\begin{table}[!t]
\centering
\small
\setlength{\tabcolsep}{3.5pt}
\renewcommand{\arraystretch}{1.08}
\begin{threeparttable}
\caption{Effect of partial leakage of \method's secret coordinate set under closed-world WBA on the FFHQ cross-family benchmark. The attacker uses surrogate detectors with guessed secrets, with the listed number of the verifier's true coordinates revealed to the surrogate. Entries are mean $\pm$ standard deviation over five runs.}
\label{tab:key-leakage}
\begin{tabular*}{\columnwidth}{@{\extracolsep{\fill}}lcc@{}}
\toprule
Revealed Coordinates & \shortstack{Removal\\ASR (\%)} & \shortstack{Forgery\\ASR (\%)} \\
\midrule
\textbf{0/32}  & \textbf{0.54}\std{0.10}  & \textbf{0.10}\std{0.03}  \\
4/32  & 0.92\std{0.22}  & 0.14\std{0.03}  \\
8/32  & 1.20\std{0.96}  & 0.18\std{0.08}  \\
16/32 & 16.42\std{4.66} & 3.18\std{2.07}  \\
24/32 & 28.62\std{4.35} & 5.50\std{3.29}  \\
32/32 & 68.90\std{6.63} & 36.88\std{5.24} \\
\bottomrule
\end{tabular*}
\begin{tablenotes}[flushleft]
\footnotesize
\item Bold indicates the lowest attack success rate in each column.
\end{tablenotes}
\end{threeparttable}
\end{table}